\documentclass[11pt]{article}
\usepackage{amssymb}
\usepackage{amsmath}
\usepackage{graphics}
\usepackage{epsfig}
\usepackage{a4wide}
\usepackage{cite}
\usepackage{multirow}
\usepackage{color}
\usepackage{mathrsfs}
\usepackage{slashed}
\usepackage{subfigure}

\DeclareGraphicsExtensions{.eps}
\begin{document}

\title{A possible interpretation for $X(6900)$ observed in four-muon final state by LHCb -- A light Higgs-like boson?}
\author{
Jian-Wen Zhu$^{1,2}$, Xing-Dao Guo$^3$\footnote{Corresponding author: guoxingdao@163.com}, Ren-You Zhang$^{1,2}$, \\
Wen-Gan Ma$^{1,2}$, and Xue-Qian Li$^4$
\\ \\
{\small $^1$ State Key Laboratory of Particle Detection and Electronics,}  \\
{\small University of Science and Technology of China, Hefei 230026, Anhui, People's Republic of China}  \\
{\small $^2$ Department of Modern Physics, University of Science and Technology of China,}  \\
{\small Hefei 230026, Anhui, People's Republic of China} \\
{\small $^3$ School of Physics and New Energy, XuZhou University of Technology,}  \\
{\small Xuzhou 221111, Jaingsu, People's Republic of China}  \\
{\small $^5$ School of Physics, Nankai University, Tianjin 300071, People's Republic of China}
}

\date{}
\maketitle
\vskip 10mm

\begin{abstract}
A peak structure  of $J/\psi$ pair production around $6.9~{\rm GeV}$ was observed and analyzed by the LHCb collaboration using the Run I and II data of LHC.
How to understand this peak arouses enthusiastic discussions among both theorists and experimentalists of high energy physics, because this discovery might
hint something new. Overwhelming works on this topic tend to attribute the peak as a four-quark state: tetraquark or molecule. Instead, we suggest that this peak is corresponding to a fundamental Higgs-like boson with mass about $6.9~{\rm GeV}$ which is advocated by a BSM effective theory. We present a detailed analysis on both signal and SM background, including integrated cross sections and invariant mass distributions of the final-state $J/\psi$ pair. Our numerical results are well in coincidence with the experimental data, as postulating  the resonance observed by LHCb to be a BSM $0^{++}$ scalar. Therefore, the peak at $M_{\text{di-}J/\psi} \sim 6.9~{\rm GeV}$ might be a hint of new physics beyond the SM whose scale is not as large as mostly expected by high energy physicists. More further works are urgently needed in both experimental and theoretical aspects to validate or negate this assumption.
\end{abstract}

\vfill \eject
\baselineskip=0.32in
\makeatletter      
\@addtoreset{equation}{section}
\makeatother       
\vskip 5mm
\renewcommand{\theequation}{\arabic{section}.\arabic{equation}}
\renewcommand{\thesection}{\Roman{section}.}
\newcommand{\nb}{\nonumber}

\section{Introduction}
\par
With the discovery of the $125~ {\rm GeV}$ Higgs boson at the Large Hadron Collider (LHC) in 2012, the last brick of the Standard Model (SM) has been laid at the designated place, the triumph of SM is unambiguously becoming the basis of describing interactions among fundamental particles. However, the SM still faces many conceptional and experimental challenges, such as the gauge hierarchy problem, the origin of neutrino masses, the dark matter and dark energy, the baryon asymmetry in our universe, etc. In a word, the SM can only be an effective theory of a larger symmetry or a more fundamental principle(s). Namely, there must exist new physics beyond the SM. Unfortunately, the experimental measurements have not provided any hints towards the new physics beyond the SM so far.
No doubt, the next goal of the updated LHC and future high-energy facilities are to precisely detect the properties of the $125~{\rm GeV}$ Higgs boson,  and search for new physics beyond the SM, concretely, one will be able to determine a scale of BSM.

\par
Since 2003, many exotic states were observed in experiments; most of them are considered as multi-quark bound states. Very recently, the LHCb collaboration analyzed the data collected during the Run I and II stages of LHC, and observed several unknown resonance structures that cannot be explained as traditional hadrons (meson and baryon). By the event reconstruction, the events with four-muon final state which is confirmed to originate from the decays of a $J/\psi$ pair are specially selected out from the database, since those $J/\psi$ can be efficiently reconstructed by a muon pair. A broad peak structure ranging from $6.2$ to $6.8~ {\rm GeV}$, a narrow peak structure at about $6.9~ {\rm GeV}$ and a hint of a peak structure at $7.2~ {\rm GeV}$ were observed in the invariant mass distribution of $J/\psi$ pair \cite{Aaij:2020fnh} in $pp$ collisions at the LHC by the LHCb collaboration. The data sets for the analysis of $J/\psi$ pair were recorded by the LHCb detector in $pp$ collisions at the center-of-mass energies of $7$, $8$ and $13~ {\rm TeV}$ with an integrated luminosity of $9~ {\rm fb}^{-1}$. The broad structure ranging from $6.2$ to $6.8~ {\rm GeV}$ is considered as a threshold enhancement, and the possible peak structure at  $7.2~ {\rm GeV}$ is excluded due to its low significance \cite{Aaij:2020fnh}. The significance of the narrow peak structure at $6.9~ {\rm GeV}$ (denoted as $X(6900)$) is $3.4\sigma$ based on unbinned maximum likelihood fits with $p_{T}^{\text{di-}J/\psi} > 5.2~ {\rm GeV}$ and $6.0\sigma$ based on maximum likelihood fits with six $p_{T}^{\text{di-}J/\psi}$ regions \cite{Aaij:2020fnh}. After considering the interference between the non-resonant single-parton scattering (NRSPS) component and a resonance for the threshold enhancement, $X(6900)$ with mass, decay width and yield of $m_X = 6.886 \pm 0.011_{\,{\rm stat.}} \pm 0.011_{\,{\rm syst.}}\, {\rm GeV}$, $\Gamma_X = 0.168 \pm 0.033_{\,{\rm stat.}} \pm 0.069_{\,{\rm syst.}}\, {\rm GeV}$ and $N_{\text{sig}}=784\pm148$ is observed in the fiducial region of \cite{Aaij:2020fnh}
\begin{equation}
\label{jpsicut}
p_T^{J/\psi} < 10.0~{\rm GeV},
\qquad\qquad
2.0 < y_{J/\psi} < 4.5.
\end{equation}

\par
From theoretical perspective, most of particle physicists believe that $X(6900)$ is a $cc\bar{c}\bar{c}$ four-quark bound state as a tetraquark, and the spectroscopy of this heavy tetraquark was studied in detail by the authors of Refs.\cite{Chen:2020xwe,Jin:2020jfc,Lu:2020cns,Yang:2020rih,Wang:2020ols,Deng:2020iqw,Chen:2020lgj,Albuquerque:2020hio,Sonnenschein:2020nwn,Giron:2020wpx,Richard:2020hdw,Becchi:2020uvq}.
However, there are still other non-resonant explanations for the peak  observed in experiments, such as the Fano-like interference mechanism \cite{Chen:2015bft,Chen:2017uof}, initial single pion emission mechanism \cite{Chen:2011pv,Chen:2011xk,Chen:2013coa,Chen:2013wca}, triangle singularity \cite{Guo:2019twa,Wang:2013hga,Wang:2013cya,Liu:2014spa,Guo:2014iya,Liu:2015taa,Nakamura:2019btl}, special three-body kinematics reflection structure \cite{Wang:2020axi}, double charmonium state rescattering\cite{Wang:2020wrp}, and so on. The peak structure, especially its peculiar decay mode enable us  to consider an alternative scenario.

\par
It is a common recognition for high energy physicists that searching for new physics should begin with searching for BSM Higgs-like boson(s). Many kinds of BSM Higgs-like bosons with different quantum numbers are predicted in various BSM models, but none of them have ever been experimentally observed so far. Generally, the thought  that the scale of new physics should be much higher than the SM Higgs mass  dominates among the high energy physicists, as it may be as high as few hundreds of GeV to few hundreds of TeV. However, this consensus is not mandatory and no any principle forbids the existence of low mass BSM particles. For example, a light Higgs-like boson with mass of $28~ {\rm GeV}$ is predicted in the Two-Higgs-Doublet Model \cite{Cici:2019zir}.

\par
Driven by the efforts of searching for BSM Higgs-like boson(s) at high-energy colliders and failure of identifying any reasonable candidate, we would think the possibility that a light Higgs-like boson might exist. Encouraging by the idea, we conjecture $X(6900)$ to be a $0^{++}$ fundamental boson. Different from the general method adopted for searching heavy particles at very high energies, we explore a possible Higgs-like boson at low energy regions in this work. As a common sense we gained from the study of lower energy experiments, such as the $e^+e^-$ collisions at BES, the processes with a resonant mediate state such as $e^+e^- \rightarrow J/\psi \rightarrow \text{{\it final products}}$ overwhelmingly dominate over the direct production which only provides the continuous background. Therefore, a direct production of $J/\psi$ pair from the gluon-gluon fusion at the LHC, $gg \rightarrow J/\psi J/\psi$\cite{Li:2009ug,Qiao:2009kg}, just compose the background to the process $gg \rightarrow X(6900) \rightarrow J/\psi J/\psi$. It is natural to conjecture that the BSM Higgs-like boson would induce a peak in the invariant mass distribution of $J/\psi$ pair. Under this assertion, we calculate the rate of $J/\psi$ pair production induced by the BSM Higgs-like boson at the LHC.

\par
The purpose of this work is to check whether this hypothesis could be tolerated by the current experimental constraints. In this paper, the production rate and invariant mass spectrum of $J/\psi$ pair at the LHC are calculated within the framework of a BSM effective theory, in which a Higgs-like boson with mass around $6.9~{\rm GeV}$ is pre-assumed. Comparing our numerical results with the signal observed by LHCb, the reasonability of the ansatz for $pp \rightarrow X(6900) \rightarrow J/\psi J/\psi \rightarrow 4\mu$ becomes more obvious. As a conclusion, our numerical results can meet the experimental data well in some specific parameter regions, and thus we would suggest that $X(6900)$ may be a BSM Higgs-like boson.

\par
This work is organized as follows. In section II, we present the analytical calculation for $J/\psi$ pair production at the LHC by taking the contribution of the BSM Higgs-like boson $X(6900)$ into account, while assuming the interactions of $X(6900)$ with gluons and quarks are in analogue to the corresponding ones of the SM Higgs boson. In section III, we provide numerical results of total cross section and yield, and illustrate the invariant mass distribution of $J/\psi$ pair. The last section is devoted to our conclusion and a brief discussion.

\section{Effective couplings to gluons and quarks}
\par
\begin{figure}[htbp]
    \centering
        \subfigure[]{
            \includegraphics[scale=0.35]{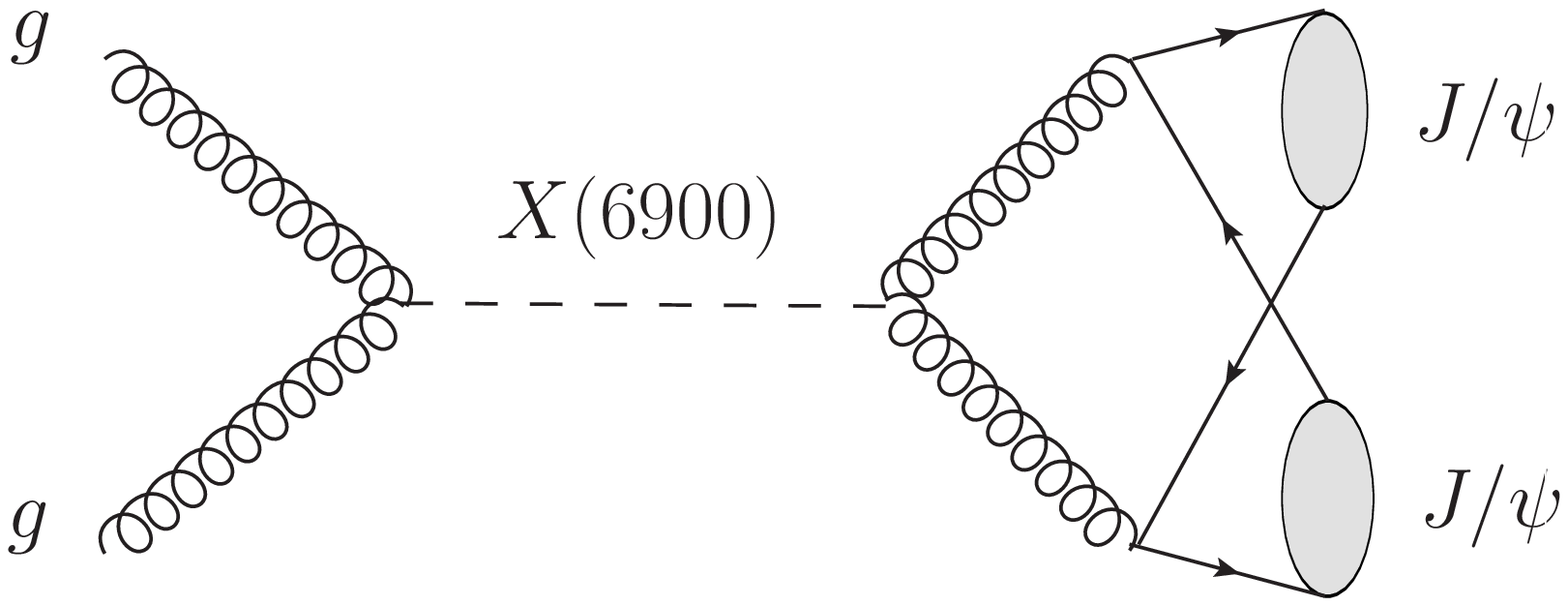}}
        \subfigure[]{
            \includegraphics[scale=0.35]{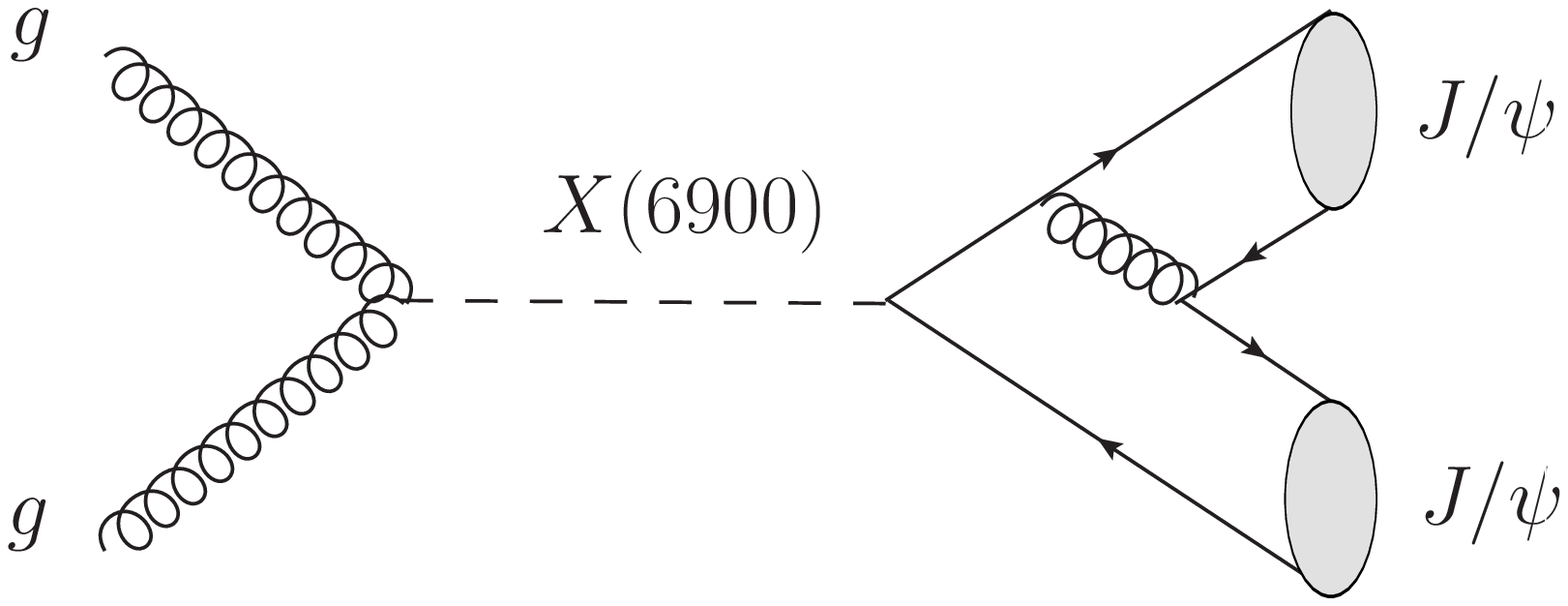}} \\
\caption{$s$-channel representative Feynman diagrams induced by $X(6900)$ for $gg \rightarrow J/\psi J/\psi$.}
\label{fig1}
\end{figure}

In this paper, we investigate in detail the $J/\psi$ pair production at the LHC within a effective theory, in which a Higgs-like scalar with mass around $3.9~ {\rm GeV}$ is introduced. At the LHC, the $J/\psi$ pair is mainly produced via gluon-gluon fusion\cite{Guo:2020vij,Georgi:1977gs,Anastasiou:2002yz}, i.e.,
\begin{equation}
\label{eqn1}
\sigma[pp \rightarrow J/\psi J/\psi]
=
\int dx_1 dx_2\, f(x_1, \mu_F)\, f(x_2, \mu_F)\, \hat{\sigma}[gg \rightarrow J/\psi J/\psi],
\end{equation}
where $f(x, \mu_F)$ is the gluon distribution function in the proton and $\mu_F$ is the factorization scale. The parton-level cross section for $gg \rightarrow J/\psi J/\psi$ is given by
\begin{equation}
\label{sigmahat}
\hat{\sigma}[gg \rightarrow J/\psi J/\psi]
=
{1\over 2\hat s}\int d \Pi_2 \, \left| \mathcal{M}_{SM} + \mathcal{M}_{X(6900)} \right|^2,
\end{equation}
where $\mathcal{M}_{SM}$ and $\mathcal{M}_{X(6900)}$ are the Feynman amplitudes in the SM and induced by the Higgs-like boson $X(6900)$, respectively. The SM Feynman amplitude $\mathcal{M}_{SM}$ can be found in Refs.\cite{Li:2009ug,Qiao:2009kg}. As for the BSM contribution from $X(6900)$, only the $ggX$ and $c\bar{c}X$ effective couplings for $X(6900)$ are considered in this paper. The $s$-channel representative Feynman diagrams induced by $X(6900)$ are shown in Fig.\ref{fig1}.

\par
Following Refs.\cite{Anastasiou:2015ema,Spira:2016ztx}, the effective couplings of the BSM Higgs-like boson $X(6900)$ to two gluons and charm quarks are expressed as
\begin{equation}
\begin{aligned}
\mathcal{C}_{ggX}^{\mu \nu}(k_1,\, k_2)
&=
-i\, \dfrac{g_{ggX}(\mu_R)}{m_X} \left[ 4k_1 \cdot k_2 \Big( g^{\mu\nu}-\frac{k_1^\nu k_2^\mu}{k_1\cdot k_2} \Big) \right],\\
\mathcal{C}_{c\bar{c}X}
&=
-i\, g_{c\bar{c}X}(\mu_R),
\end{aligned}
\end{equation}
where $g_{ggX}(\mu_R)$ and $g_{c\bar{c}X}(\mu_R)$ are dimensionless effective running coupling constants and $\mu_R$ is the renormalization scale. It is reasonable to assume that the evolution behaviors of the effective coupling constants $g_{ggX}(\mu_R)$ and $g_{c\bar{c}X}(\mu_R)$ are the same as the QCD strong coupling constant $\alpha_s(\mu_R)$ and the charm-quark $\overline{{\rm MS}}$ running mass $\overline{m}_c(\mu_R)$ \cite{Tanabashi:2018oca}, respectively. By this definition, both $\dfrac{g_{ggX}(\mu_R)}{\alpha_s(\mu_R)}$ and $\dfrac{g_{c\bar{c}X}(\mu_R)}{\overline{m}_c(\mu_R)}$ are independent of $\mu_R$. The explicit expression for $\mathcal{M}_{X(6900)}$ can be written with the help of {\sc FeynArts} \cite{Hahn:2000kx} and {\sc FeynCalc} \cite{Shtabovenko:2020gxv} packages. After those calculation, we can obtain the total cross section for $pp \rightarrow J/\psi J/\psi$ by a convolution with the gluon distribution function.

\section{Numerical results}
\par
In our calculation, the event samples are generated by using {\sc FormCalc} \cite{Hahn:2016ebn} package based on the Monte Carlo technique, and the mass and width of $X(6900)$ are set as $m_X = 6.886~{\rm GeV}$ and $\Gamma_X = 168~{\rm MeV}$ according to Ref.\cite{Aaij:2020fnh}. Within the framework of NRQCD, the zero point wave function of $J/\psi$ is needed. Following Refs.\cite{Li:2009ug,Quigg:1977dd,Eichten:1994gt,Eichten:1995ch}, we take $\Psi_{J/\psi}^2(0) = 0.064~ {\rm GeV}^3$ and $Br(J/\psi \to \mu^+\mu^-) = 5.96\%$. The factorization and renormalization scales are taken as $\mu_F = \mu_R = \sqrt{4m_c^2 + \big(p_T^{J/\psi}\big)^2}$. The masses of $c$-quark and $J/\psi$ are taken as $m_c = 1.55~ {\rm GeV}$ and $m_{J/\psi} = 3.10~ {\rm GeV}$, respectively \cite{Tanabashi:2018oca}. The gluon distribution function and the strong coupling constant $\alpha_s(\mu_R)$ are adopted from {\sc CT14LO} \cite{Schmidt:2015zda}.

\par
\begin{figure}[htbp]
\begin{center}
\includegraphics[width=0.5\textwidth]{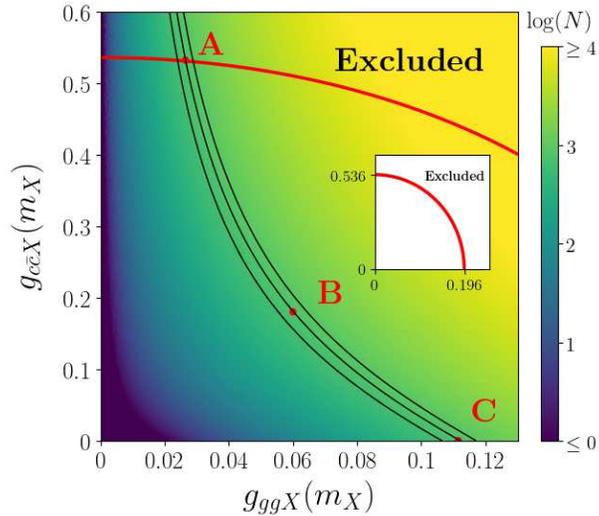}
\caption{Dependence of $N_{{\rm sig}}$ on the effective coupling constants $g_{ggX}$ and $g_{c\bar{c}X}$ at a combination of the luminosity of $\sqrt{s} = 7, 8$ and $13~{\rm TeV}$ LHC. The parameter space region above the red line is excluded by the experimental constraint of $\Gamma[X(6900) \rightarrow gg] + \Gamma[X(6900) \rightarrow c\bar{c}] < \Gamma[X(6900) \rightarrow all] \simeq 168~ {\rm MeV}$. The black lines stand for the experimental constraint given by the yield of $N_{{\rm sig}} = 784 \pm 148$.
}
\label{fig2}
\end{center}
\end{figure}

Below we present our numerical results at a combination of the luminosity of $\sqrt s=7$, $8$ and $13~{\rm TeV}$ LHC \cite{Aaij:2015awa,Aaij:2017egv}. The dependence of the yield $N_{{\rm sig}}$, defined as
\begin{equation}
N_{{\rm sig}} = \sigma_X(7~{\rm TeV})\cdot 1~{\rm fb}^{-1} + \sigma_X(8 ~{\rm TeV})\cdot 2~{\rm fb}^{-1} + \sigma_X(13 ~{\rm TeV})\cdot 6~{\rm fb}^{-1},
\end{equation}
on the effective coupling constants $g_{ggX}$ and $g_{c\bar{c}X}$ is shown in Fig.\ref{fig2}, where $\sigma_X(\sqrt{s})$ is the cross section for $pp \rightarrow J/\psi J/\psi \rightarrow \mu^+\mu^-\mu^+\mu^-$ at the LHC calculated by $\left| \mathcal{M}_{X(6900)}\right|^2$. The narrow width approximation (NWA) \cite{Kauer:2007zc,Uhlemann:2008pm} is adopted to generate the four-muon events and the following event selection criteria are applied on each final-state muon \cite{Aaij:2020fnh},
\begin{equation}
\label{mucut}
p_T^{\mu} > 0.6~{\rm GeV},
\qquad
|\vec{p}_{\mu}| > 6~{\rm GeV},
\qquad
2.0 < y_{\mu} < 4.5.
\end{equation}
In Fig.\ref{fig2}, different colors represent different values of $N_{{\rm sig}}$. The parameter space region above the red line is excluded by the experimental constraint of
\begin{equation}
\Gamma[X(6900) \rightarrow gg]+ \Gamma[X(6900) \rightarrow c\bar{c}] < \Gamma[X(6900) \rightarrow all] \simeq 168~ {\rm MeV}.
\end{equation}
The black lines stand for the experimental constraint given by the yield of $N_{{\rm sig}} = 784 \pm 148$. Thus, considering those experimental constraints mentioned above, only the black-curve band in Fig.\ref{fig2} is the experimentally allowed parameter region. In the following, we take three different benchmark points $A$, $B$ and $C$ on the central black line for comparison in our numerical calculation.

\par
\begin{figure}[htbp]
\begin{center}
\includegraphics[width=0.6\textwidth]{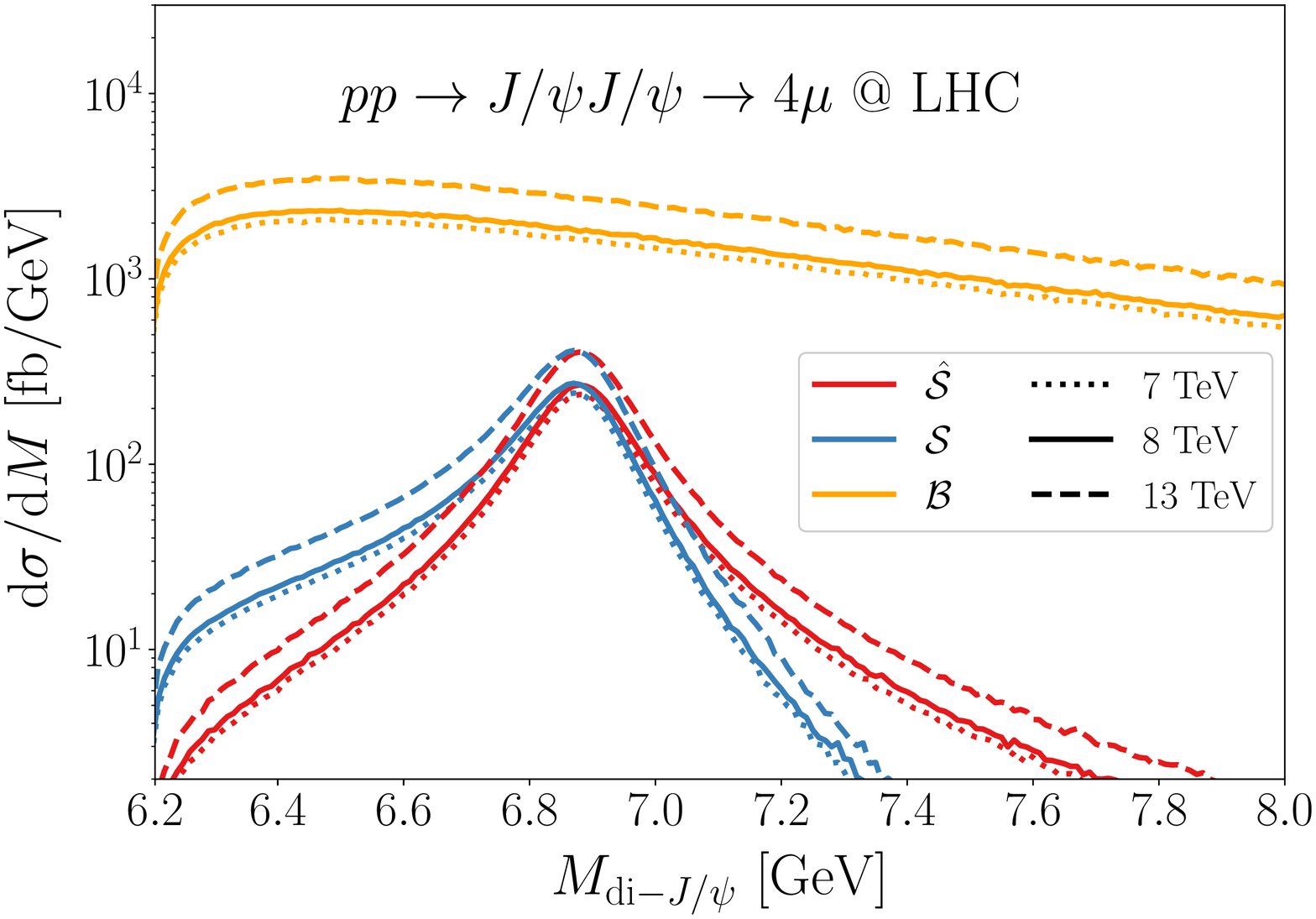}
\includegraphics[width=0.6\textwidth]{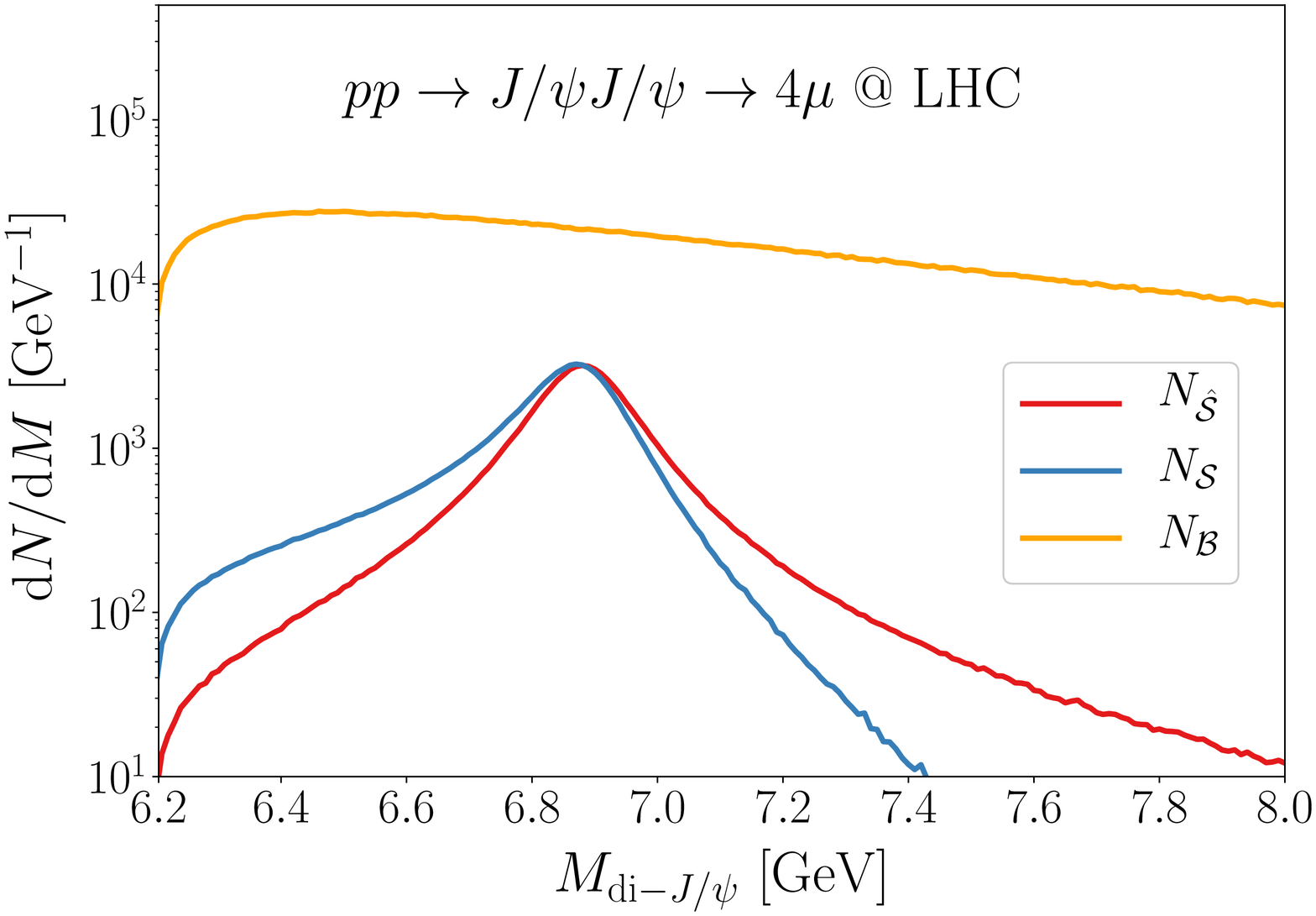}
\caption{
$\text{Di-}J/\psi$ invariant mass distribution of cross sections (upper) and event numbers (lower) for both signal and SM background for $pp \rightarrow J/\psi J/\psi \rightarrow 4\mu$ at the benchmark point $B$. $S$ and $\hat{S}$ represent the new physics signals induced by $X(6900)$ with and without interference effect, respectively.}
\label{fig3}
\end{center}
\end{figure}

The integrated cross sections and the $\text{di-}J/\psi$ invariant mass distributions for $pp \rightarrow J/\psi J/\psi \rightarrow 4\mu$ at $\sqrt s=7$, $8$ and $13 ~{\rm TeV}$ LHC at the benchmark point $B~(g_{ggX}(m_X) = 0.06,~ g_{c\bar{c}X}(m_X) = 0.18)$ are shown in Tab.\ref{tab1} and Fig.\ref{fig3}, where $B$, $S$ and $\hat{S}$ represent the SM background, the contributions from $X(6900)$ with and without interference effect, respectively. Tab.\ref{tab1} clearly shows that the contribution of the interference between $\mathcal{M}_{X(6900)}$ and $\mathcal{M}_{SM}$ is much smaller than that of $|\mathcal{M}_{X(6900)}|^2$.
We can see that our numerical results are well coincident with the experimental data, and thus support our conjecture that the resonance observed in the invariant mass distribution of $J/\psi$ pair at the LHC would be a Higgs-like boson of around $6.9~ {\rm GeV}$. Results at benchmark points $A$ and $C$ are not provided in this paper, since the signal cross sections at those two benchmark points are almost the same as that at benchmark point $B$.
\begin{table}[htbp]
\begin{center}
\renewcommand\arraystretch{1.8}
\begin{tabular}{cccc}
\hline
\hline
~~$\sqrt{s}$~ [TeV]~~ & ~~$\sigma_{\hat{S}}$~ [fb]~~ & ~~$\sigma_{S}$~ [fb]~~ & ~~$\sigma_{B}$~ [pb]~~ \\
\hline
$7$   & $58.15$ & $62.48$ & $2.96$ \\
$8$   & $65.50$ & $70.33$ & $3.34$ \\
$13$  & $98.39$ & $105.36$ & $5.05$ \\
\hline
\hline
\end{tabular}
\caption{Integrated cross sections for $pp \rightarrow J/\psi J/\psi \rightarrow 4\mu$ at benchmark point $B$ at the $7$, $8$ and $13~ {\rm TeV}$ LHC.
$B$, $S$ and $\hat{S}$ represent the SM background and the new physics signals induced by $X(6900)$ with and without interference effect, respectively.
}
\label{tab1}
\end{center}
\end{table}

\section{Discussions and conclusion}
\par
The newly observed peak structure at $6.9~ {\rm GeV}$ by the LHCb collaboration\cite{Aaij:2020fnh} hints that it may correspond to a BSM Higgs-like boson $X(6900)$, and by the anzatz we numerically calculate the cross section of the process $pp \rightarrow J/\psi J/\psi \rightarrow 4 \mu$ within a BSM effective theory in which a Higgs-like boson with mass around $6.9~ {\rm GeV}$ is introduced. In our calculation, we assume the evolution behaviors of the effective coupling constants of the Higgs-like boson to gluons and charm-quarks are the same as the corresponding ones of the SM Higgs boson.

\par
Fig.\ref{fig2} clearly shows that the region below the red line can survive as the experimental constraint is taken into account. While considering the experiment data for $pp \rightarrow J/\psi J/\psi \rightarrow 4 \mu$ collected by LHCb, the parameter space is further constrained; only the region surrounded by the black lines is experimentally allowed. From Tab.\ref{tab1} and Fig.\ref{fig3} we can find that the resonance structure can be observed clearly in the invariant mass distribution of $J/\psi$ pair with hadronic energies $\sqrt s=7$, $8$ and $13~ {\rm TeV}$, and the total cross section from the Higgs-like boson $X(6900)$ is $\mathcal{O}(10~{\rm pb})$ that can be easily measured in present facilities.

\par
Since the LHCb collaboration has announced that they have observed three resonances with a broad peak structure ranging from $6.2$ to $6.8~ {\rm GeV}$, a narrow peak structure at $6.9~ {\rm GeV}$ and a hint of peak structure at $7.2~ {\rm GeV}$\cite{Aaij:2020fnh}, much more studies on the peak should be carried out. Now the broad structure ranging from $6.2$ to $6.8~ {\rm GeV}$ is considered as a threshold enhancement and the structure at $7.2~ {\rm GeV}$ was neglected due to its low significance. Only the structure at $6.9~ {\rm GeV}$ was confirmed as a resonance.

\par
Different from the most recent researches which consider $X(6900)$ as a composed particle of four charm quarks, in this study we consider it as a BSM Higgs-like boson. By our assumption, one of the three observed peaks is a BSM Higgs-like boson, if it is true, the measurement of the LHCb collaboration would set a scale for the BSM physics and the significance is obvious. Since it implies new understanding on new physics BSM and sets a new scale, obviously, the study along this line cannot be neglected. We hope the experimentalists of high energy physics to continue the investigation on the two peaks by more accurate measurement and analysis. The conclusion would greatly help theorists making a definite judgement to verify the validity of our ansatz or negate it. All of our estimates are based on the experimental observations made by the LHCb collaboration with $\sqrt{s}$ being $7$, $8$ and $13~ {\rm TeV}$.

\vskip 10mm
\par
\noindent{\large\bf ACKNOWLEDGMENTS} \\
This work is supported in part by the National Natural Science Foundation of China (Grants Nos. 11675082 and 11735010, 11775211, 11535002, 11805160,  11747040, 11675082, 11375128,  12075125, 12035009 and the CAS Center for Excellence in Particle Physics (CCEPP).


\vskip 5mm
\begin{thebibliography}{99}

\bibitem{Aaij:2020fnh}
  R.~Aaij {\it et al.} [LHCb Collaboration],
  arXiv:2006.16957 [hep-ex].

\bibitem{Chen:2020xwe}
  H.~X.~Chen, W.~Chen, X.~Liu and S.~L.~Zhu,
  arXiv:2006.16027 [hep-ph].

\bibitem{Jin:2020jfc}
  X.~Jin, Y.~Xue, H.~Huang and J.~Ping,
  arXiv:2006.13745 [hep-ph].

\bibitem{Lu:2020cns}
  Q.~F.~L${\rm \ddot{u}}$, D.~Y.~Chen and Y.~B.~Dong,
  arXiv:2006.14445 [hep-ph].

\bibitem{Yang:2020rih}
  G.~Yang, J.~Ping, L.~He and Q.~Wang,
  arXiv:2006.13756 [hep-ph].

\bibitem{Deng:2020iqw}
  C.~Deng, H.~Chen and J.~Ping,
  arXiv:2003.05154 [hep-ph].

\bibitem{Wang:2020ols}
  Z.~G.~Wang,
  arXiv:2006.13028 [hep-ph].

\bibitem{Chen:2020lgj}
  X.~Chen,
  arXiv:2001.06755 [hep-ph].

\bibitem{Albuquerque:2020hio}
  R.~M.~Albuquerque, S.~Narison, A.~Rabemananjara, D.~Rabetiarivony and G.~Randriamanatrika,
  arXiv:2008.01569 [hep-ph].

\bibitem{Sonnenschein:2020nwn}
  J.~Sonnenschein and D.~Weissman,
  arXiv:2008.01095 [hep-ph].

\bibitem{Giron:2020wpx}
  J.~F.~Giron and R.~F.~Lebed,
  arXiv:2008.01631 [hep-ph].

\bibitem{Richard:2020hdw}
  J.~M.~Richard,
  arXiv:2008.01962 [hep-ph].

\bibitem{Becchi:2020uvq}
  C.~Becchi, A.~Giachino, L.~Maiani and E.~Santopinto,
  arXiv:2006.14388 [hep-ph].


\bibitem{Chen:2015bft}
  D.~Y.~Chen, X.~Liu, X.~Q.~Li and H.~W.~Ke,
  Phys.\ Rev.\ D {\bf 93}, 014011 (2016).

\bibitem{Chen:2017uof}
  D.~Y.~Chen, X.~Liu and T.~Matsuki,
  Eur.\ Phys.\ J.\ C {\bf 78}, 136 (2018).

\bibitem{Chen:2011pv}
  D.~Y.~Chen and X.~Liu,
  Phys.\ Rev.\ D {\bf 84}, 094003 (2011).

\bibitem{Chen:2011xk}
  D.~Y.~Chen and X.~Liu,
  Phys.\ Rev.\ D {\bf 84}, 034032 (2011).

\bibitem{Chen:2013coa}
  D.~Y.~Chen, X.~Liu and T.~Matsuki,
  Phys.\ Rev.\ D {\bf 88}, 036008 (2013).

\bibitem{Chen:2013wca}
  D.~Y.~Chen, X.~Liu and T.~Matsuki,
  Phys.\ Rev.\ Lett.\  {\bf 110}, 232001 (2013).

\bibitem{Guo:2019twa}
  F.~K.~Guo, X.~H.~Liu and S.~Sakai,
  Prog.\ Part.\ Nucl.\ Phys.\  {\bf 112}, 103757 (2020).

\bibitem{Wang:2013hga}
  Q.~Wang, C.~Hanhart and Q.~Zhao,
  Phys.\ Lett.\ B {\bf 725}, 106 (2013).

\bibitem{Wang:2013cya}
  Q.~Wang, C.~Hanhart and Q.~Zhao,
  Phys.\ Rev.\ Lett.\  {\bf 111}, 132003 (2013).

\bibitem{Liu:2014spa}
  X.~H.~Liu,
  Phys.\ Rev.\ D {\bf 90}, 074004 (2014).

\bibitem{Guo:2014iya}
  F.~K.~Guo, C.~Hanhart, Q.~Wang and Q.~Zhao,
  Phys.\ Rev.\ D {\bf 91}, 051504 (2015).

\bibitem{Liu:2015taa}
  X.~H.~Liu, M.~Oka and Q.~Zhao,
  Phys.\ Lett.\ B {\bf 753}, 297 (2016).

\bibitem{Nakamura:2019btl}
  S.~X.~Nakamura and K.~Tsushima,
  Phys.\ Rev.\ D {\bf 100}, 051502 (2019).

\bibitem{Wang:2020axi}
  J.~Z.~Wang, D.~Y.~Chen, X.~Liu and T.~Matsuki,
  arXiv:2007.02263 [hep-ph].

\bibitem{Wang:2020wrp}
  J.~Z.~Wang, D.~Y.~Chen, X.~Liu and T.~Matsuki,
  arXiv:2008.07430 [hep-ph].


\bibitem{Cici:2019zir}
  A. Cici, S. Khalil, B. Nis, and C. S. Un, arXiv:1909.02588 [hep-ph].

\bibitem{Li:2009ug}
  R. Li, Y. J. Zhang and K. T. Chao, Phys. Rev. D {\bf 80}, 014020 (2009).

\bibitem{Qiao:2009kg}
  C. F. Qiao, L. P. Sun and P. Sun, J. Phys. G {\bf 37}, 075019 (2010).

\bibitem{Guo:2020vij}
  X.~D.~Guo, J.~W.~Zhu, R.~Y.~Zhang, S.~M.~Zhao, W.~G.~Ma and X.~Q.~Li,
  arXiv:2005.04822 [hep-ph].

\bibitem{Georgi:1977gs}
  H. M. Georgi, S. L. Glashow, M. E. Machacek and D. V. Nanopoulos, Phys. Rev. Lett.  {\bf 40}, 692 (1978).

\bibitem{Anastasiou:2002yz}
  C. Anastasiou and K. Melnikov, Nucl. Phys. B {\bf 646}, 220 (2002).

\bibitem{Tanabashi:2018oca}
  M.Tanabashi {\it et al.} [Particle Data Group], Phys. Rev. D {\bf 98}, 030001 (2018).

\bibitem{Hahn:2000kx}
  T. Hahn, Comput. Phys. Commun.  {\bf 140}, 418 (2001).

\bibitem{Shtabovenko:2020gxv}
  V.~Shtabovenko, R.~Mertig and F.~Orellana,
  Comput.\ Phys.\ Commun.\  {\bf 256}, 107478 (2020)

\bibitem{Anastasiou:2015ema}
  C. Anastasiou, C. Duhr, F. Dulat, F. Herzog and B. Mistlberger, Phys. Rev. Lett.  {\bf 114}, 212001 (2015).

\bibitem{Spira:2016ztx}
  M. Spira, Prog. Part. Nucl. Phys.  {\bf 95}, 98 (2017).

\bibitem{Hao:2006nf}
  G. Hao, Y. Jia, C. F. Qiao and P. Sun, JHEP {\bf 0702}, 057 (2007).

\bibitem{Hahn:2016ebn}
  T. Hahn, S. Pa\ss ehr and C. Schappacher, PoS LL {\bf 2016}, 068 (2016), J. Phys. Conf. Ser. {\bf 762}, 012065 (2016).

\bibitem{Quigg:1977dd}
  C. Quigg and J. L. Rosner, Phys. Lett. {\bf 71B}, 153 (1977).

\bibitem{Eichten:1994gt}
  E. J. Eichten and C. Quigg, Phys. Rev. D {\bf 49}, 5845 (1994).

\bibitem{Eichten:1995ch}
  E. J. Eichten and C. Quigg, Phys. Rev. D {\bf 52}, 1726 (1995).

\bibitem{Schmidt:2015zda}
  C. Schmidt, J. Pumplin, D. Stump and C. P. Yuan, Phys. Rev. D {\bf 93}, 114015 (2016).

\bibitem{Aaij:2015awa}
R.~Aaij et al. [LHCb],
JHEP {\bf 11}, 103 (2015)

\bibitem{Aaij:2017egv}
R.~Aaij et al. [LHCb],
JHEP {\bf 12}, 110 (2017)

\bibitem{Kauer:2007zc}
  N.~Kauer,
  Phys.\ Lett.\ B {\bf 649}, 413 (2007)

\bibitem{Uhlemann:2008pm}
  C.~F.~Uhlemann and N.~Kauer,
  Nucl.\ Phys.\ B {\bf 814}, 195 (2009)



\end{thebibliography}
\end{document}